 \documentclass[final,authoryear,5p]{elsarticle}

\usepackage{epsfig}

\usepackage{amssymb}

\usepackage[ps2pdf,%
a4paper=true,%
breaklinks=true,%
colorlinks=true,%
pdfauthor={First Author et al.},%
pdftitle={Template for manuscripts in Advances in Space Research}%
]{hyperref}

\journal{Advances in Space Research}

\begin{document}

\begin{frontmatter}



\title{Stark broadening of B IV lines for astrophysical and laboratory
plasma research\tnoteref{footnote1}}
\tnotetext[footnote1]{This template can be used for all publications in Advances in Space Research.}


\author[label1,label2]{Milan S. Dimitrijevi\'c}
\ead{mdimitrijevic@aob.rs}
\address[label1]{Astronomical Observatory, Volgina 7, 11060 Belgrade 38, Serbia.}
\address[label2]{Laboratoire d'\'{E}tude du Rayonnement et de la
Mati\`{e}re en Astrophysique,Observatoire de Paris,\\
UMR CNRS 8112, UPMC, 5 Place Jules Janssen, 92195 Meudon Cedex, France.}


\author[label3]{Magdalena Christova}
\ead{mchristo@tu-sofia.bg}
\address[label3]{Department of Applied Physics, Technical University-Sofia, 1000 Sofia, Bulgaria}

\author[label1]{Zoran Simi\'c}
\ead{zsimic@aob.bg.ac.rs}

\author[label4]{Andjelka \-  Kova\v cevi\'c}
\ead{andjelka@matf.bg.ac.rs}
\address[label4]{Faculty of Mathematics, University of Belgrade, 11000 Belgrade, Serbia}

\author[label2]{Sylvie Sahal-Br\'echot}
\ead{sylvie.sahal-brechot@obspm.fr}

\begin{abstract}

Stark broadening parameters for 36 multiplets of B IV have been calculated using
the semi-classical perturbation formalism. Obtained results have been used to investigate the regularities
within spectral series and temperature dependence.

\end{abstract}


\end{frontmatter}

\parindent=0.5 cm

\section{Introduction}

The study presents Stark broadening parameters (widths and shifts) of B IV spectral lines
which have been determined using the semi-classical perturbation formalism \citep{Sah69a,Sah69b}.
Data on boron lines are of interest in astrophysics, astrochemistry, and cosmology,
in technological plasma research, for thermonuclear reaction devices, and for laser produced plasma investigations.
For abundance determinations of boron and modelling, and analysis of stellar plasma it is necessary to have
reliable atomic and spectroscopic data, including Stark broadening parameters. This enables to provide data
on the astrophysical processes that can both produce and destroy this rare element. Namely, the light elements
lithium, beryllium, and boron (LiBeB) are sensitive probes of stellar models due to the fact that the stable isotopes
of all three consist of nuclei with small binding energies that are destroyed easily by (p, a) reactions at modest
temperatures \citep{CS99}. The origin and evolution of boron are of special interest because it is hardly
produced by the standard big bang nucleosynthesis (BBN), and cannot be produced by nuclear fusions in stellar interiors
\citep{Tan03}. The cosmic abundance of $^{11}$B is of major importance for the model of Galactic chemical evolution
(GCE) \citep{Rit11}. \citet{Rit11} concluded that a major portion of the cosmic abundance of $^{11}$B can be
attributed to neutrino nucleosynthesis. Thus, it is necessary to accurately describe the stellar evolution, and the formation
of elements, which are closely connected. To make progress in these developments chemical
abundances are crucial parameters to be determined. This needs an accurate interpretation of the detailed
line spectra of the stellar objects.
In order to provide the data needed in astrophysics, laboratory-, technological-, fusion-, and laser produced-plasma research,
our aim is to determine here Stark broadening parameters (full widths at half intensity and shifts) for 36 B IV multiplets.
Obtained results will be also used for the consideration of regulariries within spectral series and temperature dependence
of Stark broadening parameters.

\section{Theory of Stark broadening in the impact approximation}

Pressure broadening of spectral lines arises when an atom, ion, or molecule which emits or absorbs light
in a gas or plasma, is perturbed by its interactions with the other particles of the medium. Interpretation of this
phenomenon is currently used for modelling of the medium and for spectroscopic diagnostics, since the broadening of
the lines depends on the temperature and density of the medium.  The physical conditions in the Universe are very various,
and collisional broadening with charged particles (Stark broadening) appears to be important in many domains. For example,
at temperatures  around 10$^4$ K and densities 10$^{13}$ - 10$^{15}$ cm$^{-3}$,
Stark broadening is of interest for modelling and analysing
spectra of  A and B type stars: see e.g. \citet{Lan88}, \citet{Pop01a,Pop01b,Pop99a,Pop99b}, \citet{Tan03},
\citet{Sim05}, \citet{Sah10}. Especially in white dwarfs, Stark broadening is the dominant
collisional line broadening mechanism in all important layers of the atmosphere
\citep{Pop99b,Tan03,Sim06,Ham08,Sim09,Dim11,Duf11,Lar12}.
The theory of Stark broadening is well applied, especially for accurate spectroscopic diagnostics
and modelling. This requires the knowledge of numerous profiles, especially for trace elements, as boron in this case,
which are used as useful probes for modern spectroscopic diagnostics. Interpretation of the spectra of white dwarfs,
which are very faint, allows understanding the evolution of these very old stars, which are close to death.
The results for Stark broadening parameters of 36 BIV multiplets have been calculated using the semi-classical
perturbation formalism \citep{Sah69a,Sah69b}. Within this theory the full half width (W) and the shift (d) of
an isolated  line originating from the transition between the initial level i and the final level f is expressed as:

  \begin{equation}
    \label{eq:1}
 {
\!\!\! \!\! \!\! \!\! \!\! \!\!\!W = n_e\int_{0}^{+\infty} vf(v)dv
[\sum_{i'\ne
i}\sigma_{ii'}(v) + \sum_{f'\ne f}\sigma_{ff'}(v) + \sigma_{el}],
    }
  \end{equation}

  \begin{equation}
    \label{eq:2}
\!\!\!\!\!\!\!\!\!\!\!d = n_e\int_{0}^{+\infty}vf(v)dv\int_{R_3}^{R_D} 2\pi \rho d\rho \sin 2\phi_p
,
\end{equation}

  \noindent where $i'$ and $f'$ are perturbing levels, $n_{e}$ and $v$ are the electron density and the velocity of perturbers respectively,
  and $f(v)$ is the Maxwellian distribution of electron velocities.

  The inelastic cross sections $\sigma_{ii'}(v)$ (respec­tively $\sigma_{ff'}(v)$) can be expressed by an integration of the
  transition probability $Pii'$ over the impact parameter $\rho$:

\begin{equation}
  \label{eq:3}
{\sum_{i'{{\not}=}i}{\sigma}_{ii'}(v)=\frac{1}{2}\pi{R_{1}}^{2}+
{\int_{R_{1}}^{R_{D}}2{\pi}{\rho}d{\rho}\sum_{i'{{\not}=}i}P_{ii'}({\rho},v)}}.
\end{equation}

The elastic collision contribution to the width is given by:

\begin{equation}
  \label{eq:4}
{{\sigma}_{el}=2\pi{R_{2}}^{2}+{\int_{R_{2}}^{R_{D}}8{\pi}{\rho}d{\rho}{\sin}^{2}
{\delta}} + \sigma_{r}},
\end{equation}

\begin{equation}
  \label{eq:5}
{{\delta}=({\phi}^{2}_{p}+{\phi}^{2}_{q})^{1/2}}.
\end{equation}

  The phase shifts ${\phi}_{p}$ and ${\phi}_{q}$ are due to the polarization and quadruple potential respectively.
  The cut-off parameters $R_1, R_2, R_3$, the Debye cut-off $R_{D}$ and the symmetrization procedure are described
  in the above mentioned papers.The contribution of Feshbach resonances, $\sigma_{r}$ is described in \citet{Fleu77}. In the following, the collisions of emitters with electrons, protons and ionized helium are examined, and the contribution of different perturbers in total Stark broadening parameters are discussed.

\section{Results and discussion}
In this paper, Stark broadening widths (FWHM) and shifts for 36 BIV multiplets have been calculated using the
semiclassical perturbation formalism \citep{Sah69a,Sah69b} and presented in Table 1. Since the Stark broadening
widths and shifts are the same for spectral lines within a multiplet when expressed in angular frequency units for a simple spectrum like the B IV one, it is easy to obtain their values in Angstr\"oms or nanometers (see e.g. Eqs. 7
and 8 in \citet{Ham13}).  In particular, the spectral line broadening due to interactions of the emitters with electrons, protons and singly charged helium ions as perturbers has been examined.

Calculations have been performed using experimental
values of energy levels in  \citet{Kra08}. The oscillator strengths have been calculated
using the Coulomb approximation method \citet{BD49} and the tables of \citet{OS68}, while for higher levels, the method
described by \citet{van79} has been applied.   The asterisks in Table 1 indicate that the impact approximation
reaches the limit of validity.

The temperature dependences of Stark widths and shifts for the
$1s2p ^{1}P^{o} - 1s3s ^{1}S, 1s2p ^{1}P^{o} - 1s3d ^{1}D, 1s3p ^{1}P^{o} - 1s4s ^{1}S$ and $1s3p ^{1}P^{o} - 1s4d ^{1}D$
transitions at an electron density of 10$^{17}$ cm$^{-3}$ have been presented in Fig. 1 and Fig. 2, respectively.
The first two transitions have the same lower energy level $2p$ where both, the width and shift of $2p - 3s$ are
larger than the Stark parameters of $2p - 3d$. For the other  transitions with the same $3p$ lower energy level,
$3p - 4d$ transition has larger width, while $3p -4s$ has larger shift. For all of them, the width and shift due to
electrons decrease slowly for temperature values above 100 000 K.

In Fig. 3 and Fig. 4, the temperature dependences of proton-impact widths and shifts, respectively, are shown.
One can see that both widths and shifts slowly increase with the temperature. For proton-impact broadening for
B IV  $3p - 4d$ transition the impact approximation is not valid at an electron density of $10^{17}$ cm$^{-3}$ within the whole
interval of the considered temperatures and they are not included in the figures.

The dependence of the broadening parameters of spectral lines due to impacts with charged particles versus
principal quantum number within a spectral series is an important information. If we know the trend of Stark
broadening parameters within a spectral series, it is possible to interpolate or extrapolate the eventually missing
values within the considered series. The regular behavior of Stark broadening widths and shifts versus principal
quantum number within spectral series is presented in Fig. 5 and Fig. 6. The both parameters increase with principal
quantum number which reflects the decrease of the distance to the nearest perturbing levels with the increase of the
principal quantum number.

\begin{onecolumn}

\begin{table}
\caption{Stark broadening parameters for singlet B IV multiplets for a perturper density of $10^{17}$ cm$^{-3}$ and temperatures
from 20 000 to 500 000 K. The width (FWHM) W and shift d (a positive shift is towards the red) values from different perturbers are given
in (\AA) for 36 multiplets. The ratio of the included parameter C versus the corresponding Stark width gives
an estimate of  the maximal pertuber density for which the line may be treated as isolated. The asterisks indicate that the impact approximation reaches the limit of validity.
W$_{e}$ - electron-impact width, d$_{e}$ - electron-impact shift, W$_{p}$ - proton-impact width, d$_{p}$ - proton-impact shift,
W$_{He+}$ - singly charged helium ion-impact width, d$_{He+}$ - singly charged helium ion-impact shift.}

\footnotesize\scriptsize
\begin{tabular}{llllllll}
\hline
Transition & T(K) & W$_{e}$(\AA) & d$_{e}$(\AA) & W$_{p}$(\AA) & d$_{p}$(\AA) & W$_{He+}$(\AA) & d$_{He+}$(\AA) \\
\hline
B IV 2p-3d  &  20000. & 0.244E-02  & 0.677E-04  & 0.735E-04  & 0.159E-03 & 0.898E-04 & 0.147E-03\\
  418.7 \AA &  50000. & 0.160E-02  & 0.111E-03  & 0.209E-03  & 0.275E-03 & 0.204E-03 & 0.241E-03\\
C= 0.24E+18 & 100000. & 0.122E-02  & 0.117E-03  & 0.348E-03  & 0.349E-03 & 0.297E-03 & 0.295E-03\\
            & 200000. & 0.953E-03  & 0.106E-03  & 0.487E-03  & 0.421E-03 & 0.387E-03 & 0.353E-03\\
            & 300000. & 0.832E-03  & 0.100E-03  & 0.586E-03  & 0.468E-03 & 0.455E-03 & 0.390E-03\\
            & 500000. & 0.710E-03  & 0.849E-04  & 0.739E-03  & 0.522E-03 & 0.550E-03 & 0.430E-03\\
\hline
B IV 2p-4d  &  20000. & 0.581E-02  & 0.221E-03  & & & & \\
  308.4 \AA &  50000. & 0.451E-02  & 0.308E-03  & & & & \\
 C= 0.17E+16& 100000. & 0.369E-02  & 0.288E-03  & & & & \\
            & 200000. & 0.298E-02  & 0.365E-03  & & & & \\
            & 300000. & 0.262E-02  & 0.303E-03  & & & & \\
            & 500000. & 0.222E-02  & 0.222E-03  & & & & \\
\hline
B IV 2p-5d  &  20000. & 0.130E-01  & 0.506E-03  & & & & \\
  274.9 \AA &  50000. & 0.104E-01  & 0.616E-03  & & & & \\
 C= 0.97E+15& 100000. & 0.856E-02  & 0.585E-03  & & & & \\
            & 200000. & 0.692E-02  & 0.854E-03  & & & & \\
            & 300000. & 0.607E-02  & 0.665E-03  & & & & \\
            & 500000. & 0.512E-02  & 0.457E-03  & & & & \\
\hline
B IV 2p-6d  &  20000. & 0.251E-01  & 0.804E-03  & & & & \\
  259.6 \AA &  50000. & 0.203E-01  & 0.102E-02  & & & & \\
 C= 0.62E+15& 100000. & 0.168E-01  & 0.985E-03  & & & & \\
            & 200000. & 0.136E-01  & 0.165E-02  & & & & \\
            & 300000. & 0.119E-01  & 0.121E-02  & & & & \\
            & 500000. & 0.101E-01  & 0.806E-03  & & & & \\
\hline
B IV 2p-7d  &  20000. &*0.411E-01  &*0.111E-02  & & & & \\
  251.2 \AA &  50000. & 0.344E-01  & 0.133E-02  & & & & \\
 C= 0.20E+15& 100000. & 0.290E-01  & 0.139E-02  & & & & \\
            & 200000. & 0.237E-01  & 0.223E-02  & & & & \\
            & 300000. & 0.209E-01  & 0.165E-02  & & & & \\
            & 500000. & 0.177E-01  & 0.115E-02  & & & & \\
\hline
B IV 3p-4d  &  20000. & 0.116      & 0.607E-02  & & & & \\
 1190.5 \AA &  50000. & 0.889E-01  & 0.723E-02  & & & & \\
 C= 0.25E+17& 100000. & 0.727E-01  & 0.677E-02  & & & & \\
            & 200000. & 0.591E-01  & 0.761E-02  & & & & \\
            & 300000. & 0.521E-01  & 0.636E-02  & & & & \\
\hline
B IV 3p-5d  &  20000. & 0.126      & 0.514E-02  & & & & \\
  809.7 \AA &  50000. & 0.999E-01  & 0.657E-02  & & & & \\
 C= 0.84E+16& 100000. & 0.822E-01  & 0.622E-02  & & & & \\
            & 200000. & 0.666E-01  & 0.841E-02  & & & & \\
            & 300000. & 0.585E-01  & 0.662E-02  & & & & \\
            & 500000. & 0.495E-01  & 0.469E-02  & & & & \\
\hline
B IV 3p-6d  &  20000. &  0.186     & 0.623E-02  & & & & \\
  689.8 \AA &  50000. &  0.150     & 0.808E-02  & & & & \\
 C= 0.44E+16& 100000. &  0.124     & 0.779E-02  & & & & \\
            & 200000. &  0.101     & 0.124E-01  & & & & \\
            & 300000. &  0.886E-01 & 0.917E-02  & & & & \\
            & 500000. &  0.749E-01 & 0.621E-02  & & & & \\
\hline
B IV 3p-7d  &  20000. & *0.269     &*0.758E-02  & & & & \\
  633.2 \AA &  50000. &  0.224     & 0.928E-02  & & & & \\
 C= 0.12E+16& 100000. &  0.189     & 0.953E-02  & & & & \\
            & 200000. &  0.155     & 0.148E-01  & & & & \\
            & 300000. &  0.137     & 0.110E-01  & & & & \\
            & 500000. &  0.116     & 0.776E-02  & & & & \\
\hline
B IV 4p-5d  & 20000.  &  1.50      & 0.745E-01  & & & & \\
 2567.3 \AA & 50000.  &  1.18      & 0.906E-01  & & & & \\
 C= 0.84E+17& 100000. &  0.973     & 0.846E-01  & & & & \\
            & 200000. &  0.789     & 0.104      & & & & \\
            & 300000. &  0.693     & 0.840E-01  & & & & \\
            & 500000. &  0.587     & 0.610E-01  & & & & \\
\hline
B IV 4p-6d  &  20000. &  1.17      & 0.455E-01  & & & & \\
 1655.2 \AA &  50000. &  0.938     & 0.568E-01  & & & & \\
 C= 0.25E+17& 100000. &  0.775     & 0.541E-01  & & & & \\
            & 200000. &  0.629     & 0.793E-01  & & & & \\
            & 300000. &  0.552     & 0.600E-01  & & & & \\
            & 500000. &  0.467     & 0.416E-01  & & & & \\
\hline
B IV 4p-7d  &  20000. & *1.31      & *0.419E-01 & & & & \\
 1363.1 \AA &  50000. &  1.09      & 0.500E-01  & & & & \\
 C= 0.58E+16& 100000. &  0.915     & 0.504E-01  & & & & \\
            & 200000. &  0.749     & 0.741E-01  & & & & \\
            & 300000. &  0.661     & 0.561E-01  & & & & \\
            & 500000. &  0.560     & 0.399E-01  & & & & \\

\hline\normalsize
\end{tabular}
\label{table1}
\end{table}

\addtocounter{table}{-1}
\begin{table}
\caption{Continued}

\footnotesize\scriptsize
\begin{tabular}{llllllll}
\hline
Transition & T(K) & W$_{e}$(\AA) & d$_{e}$(\AA) & W$_{p}$(\AA) & d$_{p}$(\AA) & W$_{He+}$(\AA) & d$_{He+}$(\AA) \\
\hline
B IV 3d-4p  &  20000. &  0.105     & -0.806E-02  &0.122E-01  &-0.128E-01  &*0.115E-01   &*-0.104E-01\\
 1163.5 \AA &  50000. &  0.768E-01 & -0.845E-02  &0.193E-01  &-0.181E-01  &*0.163E-01   &*-0.151E-01\\
 C= 0.76E+18& 100000. & 0.615E-01  & -0.783E-02  &0.255E-01  &-0.222E-01  &*0.205E-01   &*-0.184E-01\\
            & 200000. & 0.496E-01  & -0.691E-02  &0.324E-01  &-0.252E-01  & 0.251E-01   &-0.210E-01\\
            & 300000. & 0.438E-01  & -0.613E-02  &0.377E-01  &-0.271E-01  & 0.281E-01   &-0.226E-01\\
            & 500000. & 0.375E-01  & -0.500E-02  &0.448E-01  &-0.300E-01  & 0.330E-01   &-0.254E-01\\
\hline
B IV 3d-5p  & 20000.  & 0.121      &-0.989E-02   & & & & \\
   798.8 \AA& 50000.  & 0.911E-01  &-0.103E-01   &*0.302E-01 &-0.265E-01 & & \\
 C= 0.18E+18& 100000. & 0.737E-01  &-0.921E-02   &*0.391E-01 &-0.314E-01 & & \\
            & 200000. & 0.596E-01  &-0.881E-02   &*0.474E-01 &-0.357E-01 & *0.367E-01 &-0.305E-01\\
            & 300000. & 0.526E-01  &-0.749E-02   &*0.524E-01 &-0.394E-01 & *0.404E-01 &-0.315E-01\\
            & 500000. & 0.448E-01  &-0.601E-02   &*0.662E-01 &-0.429E-01 & *0.466E-01 &-0.361E-01\\
\hline
B IV 3d-6p  &  20000. & 0.174     &  -0.138E-01  & & & & \\
   682.5 \AA&  50000. & 0.137     &  -0.146E-01  & & & & \\
 C= 0.77E+17& 100000. & 0.113     &  -0.133E-01  & & & & \\
            & 200000. & 0.925E-01 &  -0.140E-01  & & & & \\
            & 300000. & 0.817E-01 &  -0.118E-01  & & & & \\
            & 500000. & 0.697E-01 &  -0.923E-02  & & & & \\
\hline
B IV 3d-7p  &  20000. &*0.257     & *-0.183E-01  & & & & \\
   627.3 \AA&  50000. & 0.210     &  -0.197E-01  & & & & \\
 C= 0.40E+17& 100000. & 0.177     &  -0.191E-01  & & & & \\
            & 200000. & 0.146     &  -0.224E-01  & & & & \\
            & 300000. & 0.129     &  -0.187E-01  & & & & \\
            & 500000. & 0.110     &  -0.141E-01  & & & & \\
\hline
B IV 4d-6p  &  20000. & 1.11      & -0.843E-01  & & & & \\
  1635.6 \AA&  50000. & 0.877     & -0.907E-01  & & & & \\
 C= 0.47E+17& 100000. & 0.727     & -0.828E-01  & & & & \\
            & 200000. & 0.594     & -0.893E-01  & & & & \\
            & 300000. & 0.524     & -0.748E-01  & & & & \\
            & 500000. & 0.446     & -0.580E-01  & & & & \\
\hline
B IV 4d-7p  &  20000. & *1.27     &*-0.883E-01  & & & & \\
  1350.9 \AA&  50000. &  1.04     & -0.962E-01  & & & & \\
 C= 0.32E+17& 100000. & 0.873     & -0.929E-01  & & & & \\
            & 200000. & 0.719     & -0.110      & & & & \\
            & 300000. & 0.637     & -0.917E-01  & & & & \\
            & 500000. & 0.544     & -0.686E-01  & & & & \\
\hline
B IV 5d-6p  &  20000. & *11.0     & *-0.749     & & & & \\
  4623.4 \AA&  50000. &  8.80     & -0.830      & & & & \\
 C= 0.27E+18& 100000. &  7.30     & -0.762      & & & & \\
            & 200000. &  5.95     & -0.873      & & & & \\
            & 300000. &  5.24     & -0.718      & & & & \\
            & 500000. &  4.45     & -0.543      & & & & \\
\hline
B IV 5d-7p  & 20000.  & *6.68     &*-0.436      & & & & \\
  2897.7 \AA& 50000.  &  5.46     & -0.484      & & & & \\
 C= 0.11E+18& 100000. &  4.60     & -0.467      & & & & \\
            & 200000. &  3.78     & -0.568      & & & & \\
            & 300000. &  3.34     & -0.469      & & & & \\
            & 500000. &  2.84     & -0.347      & & & & \\
\hline
B IV 3d-4f  &  20000. &  0.522E-01 &  0.145E-03  & & & & \\
  1170.9 \AA&  50000. &  0.395E-01 & -0.224E-03  & & & & \\
 C= 0.24E+17& 100000. &  0.319E-01 & -0.401E-03  & & & & \\
            & 200000. &  0.258E-01 & -0.114E-02  & & & & \\
            & 300000. &  0.228E-01 & -0.793E-03  & & & & \\
            & 500000. &  0.196E-01 & -0.228E-03  & & & & \\
\hline
B IV 3d-5f  &  20000. &  0.817E-01 & 0.691E-03   & & & & \\
   800.6 \AA&  50000. &  0.679E-01 & 0.109E-03   & & & & \\
 C= 0.21E+16& 100000. &  0.568E-01 & 0.175E-03   & & & & \\
            & 200000. &  0.464E-01 &-0.965E-03   & & & & \\
            & 300000. &  0.409E-01 &-0.573E-03   & & & & \\
            & 500000. &  0.347E-01 & 0.297E-04   & & & & \\
\hline
B IV 3d-6f  &  20000.   & 0.140      & 0.139E-02   & & & & \\
   683.2 \AA&  50000.   & 0.119      & 0.553E-03   & & & & \\
 C= 0.10E+16& 100000.   & 0.996E-01  & 0.536E-03   & & & & \\
            & 200000.   & 0.813E-01  &-0.148E-02   & & & & \\
            & 300000.   & 0.716E-01  &-0.640E-03   & & & & \\
            & 500000.   & 0.607E-01  & 0.244E-03   & & & & \\
\hline
B IV 3d-7f  &  20000.  & *0.220     &*0.170E-02    & & & & \\
   627.7 \AA&  50000.  &  0.190     & 0.148E-02    & & & & \\
 C= 0.67E+15& 100000.  &  0.162     & 0.133E-02    & & & & \\
            & 200000.  &  0.133     & 0.301E-03    & & & & \\
            & 300000.  &  0.118     & 0.872E-03    & & & & \\
            & 500000.  &  0.998E-01 & 0.103E-02    & & & & \\

\hline\normalsize
\end{tabular}
\label{table1}
\end{table}

\addtocounter{table}{-1}
\begin{table}
\caption{Continued}

\footnotesize\scriptsize
\begin{tabular}{llllllll}
\hline
Transition & T(K) & W$_{e}$(\AA) & d$_{e}$(\AA) & W$_{p}$(\AA) & d$_{p}$(\AA) & W$_{He+}$(\AA) & d$_{He+}$(\AA) \\

\hline
B IV 4d-5f  &  20000.  &  1.08     & -0.507E-02    & & & & \\
  2530.1 \AA&  50000.  &  0.900    & -0.156E-01    & & & & \\
 C= 0.21E+17& 100000.  &  0.753    & -0.134E-01    & & & & \\
            & 200000.  &  0.615    & -0.304E-01    & & & & \\
            & 300000.  &  0.542    & -0.225E-01    & & & & \\
            & 500000.  &  0.460    & -0.116E-01    & & & & \\
\hline
B IV 4d-6f  &  20000.  & 0.920     &  0.264E-02    & & & & \\
  1639.7 \AA&  50000.  & 0.775     & -0.399E-02    & & & & \\
 C= 0.59E+16& 100000.  & 0.651     & -0.334E-02    & & & & \\
            & 200000.  & 0.532     & -0.173E-01    & & & & \\
            & 300000.  & 0.468     & -0.107E-01    & & & & \\
            & 500000.  & 0.396     & -0.358E-02    & & & & \\
\hline
B IV 4d-7f  &  20000.  &*1.10      & *0.394E-02    & & & & \\
  1352.8 \AA& 50000.   & 0.943     &  0.187E-02    & & & & \\
 C= 0.31E+16& 100000.  & 0.802     &  0.170E-02    & & & & \\
            & 200000.  & 0.661     & -0.457E-02    & & & & \\
            & 300000.  & 0.583     & -0.760E-03    & & & & \\
            & 500000.  & 0.495     &  0.137E-02    & & & & \\
\hline
B IV 5d-6f  &  20000.  & 9.65      & -0.507E-01    & & & & \\
  4657.0 \AA&  50000.  & 8.08      & -0.137        & & & & \\
 C= 0.48E+17& 100000.  & 6.77      & -0.128        & & & & \\
            & 200000.  & 5.52      & -0.301        & & & & \\
            & 300000.  & 4.85      & -0.208        & & & & \\
            & 500000.  & 4.10      & -0.109        & & & & \\
\hline
B IV 5d-7f  &  20000.  &*5.91      &*-0.114E-01    & & & & \\
  2906.0 \AA&  50000.  & 5.06      & -0.328E-01    & & & & \\
 C= 0.14E+17& 100000.  & 4.29      & -0.320E-01    & & & & \\
            & 200000.  & 3.52      & -0.841E-01    & & & & \\
            & 300000.  & 3.10      & -0.509E-01    & & & & \\
            & 500000.  & 2.63      & -0.250E-01    & & & & \\
\hline
B IV 4f-5d  &  20000.  & 1.23      &  0.315E-01    & & & & \\
  2532.0 \AA&  50000.  & 0.990     &  0.483E-01    & & & & \\
 C= 0.82E+17& 100000.  & 0.817     &  0.468E-01    & & & & \\
            & 200000.  & 0.662     &  0.737E-01    & & & & \\
            & 300000.  & 0.582     &  0.563E-01    & & & & \\
            & 500000.  & 0.492     &  0.367E-01   & & & & \\
 \hline
B IV 4f-6d  & 20000.   & 1.05    &    0.262E-01    & & & & \\
  1640.5 \AA& 50000.   & 0.855   &    0.385E-01    & & & & \\
 C= 0.25E+17& 100000.  & 0.708   &    0.380E-01    & & & & \\
            & 200000.  & 0.574   &    0.663E-01    & & & & \\
            & 300000.  & 0.505   &    0.483E-01    & & & & \\
            & 500000.  & 0.426   &    0.313E-01    & & & & \\
\hline
B IV 4f-7d  &  20000.  & *1.23   &    *0.276E-01   & & & & \\
  1353.1 \AA& 50000.   & 1.03    &    0.372E-01    & & & & \\
 C= 0.57E+16& 100000.  & 0.866   &    0.392E-01    & & & & \\
            & 200000.  & 0.710   &    0.650E-01    & & & & \\
            & 300000.  & 0.626   &    0.479E-01    & & & & \\
            & 500000.  & 0.530   &    0.328E-01    & & & & \\
\hline
B IV 5f-6d  & 20000.   & *10.3   &    *0.189       & & & & \\
  4661.8 \AA& 50000.   & 8.45    &    0.304        & & & & \\
 C= 0.72E+17& 100000.  & 7.04    &    0.294        & & & & \\
            & 200000.  & 5.72    &    0.550        & & & & \\
            & 300000.  & 5.03    &    0.397        & & & & \\
            & 500000.  & 4.25    &    0.248        & & & & \\
\hline
B IV 5f-7d  & 20000.   & *6.34   &    *0.117       & & & & \\
  2907.4 \AA& 50000.   & 5.34    &    0.168        & & & & \\
 C= 0.26E+17& 100000.  & 4.51    &    0.176        & & & & \\
            & 200000.  & 3.69    &    0.306        & & & & \\
            & 300000.  & 3.26    &    0.224        & & & & \\
            & 500000.  & 2.76    &    0.150        & & & & \\
\hline\normalsize
\end{tabular}
\label{table1}
\end{table}
\end{onecolumn}

\begin{twocolumn}

\section{Conclusion}

New Stark broadening parameters for 36 multiplets of B IV have been determined within the frame of the
semi-classical perturbation formalism.The results for Stark broadening parameters of boron lines could be
applicable for the adequate interpretation of the observed spectra in astrophysics, astrochemistry, and cosmology,
in technological plasma research, for thermonuclear reaction devices, and for laser produced plasma investigation.
We also note that the Stark broadening data obtained in the present research, will be inserted in the STARK-B database
\citep{Sah12,Sah13}, which is a part of Virtual Atomic and Molecular Data Center
(VAMDC - \citep{Dub10, Rix10}).

Additionally, it has been confirmed that the behaviour of Stark broadening parameters within B IV spectral series
is regular, enabling the interpolation and extrapolation of new data.

\section*{Acknowledgments}
This paper is within the projects 176002 and III44022 of Ministry of Education, Science and Technological Development
of Republic of Serbia. It is also the result of  Short term Scientific Mission CM0805 within COST programme on
The Chemical Cosmos: Understanding Chemistry in Astronomical Environments. Partial financial support from project No
6572-20 Technical University - Sofia within Euroatom Programme is also acknowledged. One of us (MSD) acknowledges the support of the LABEX PLAS@PAR.


\begin{figure}
\label{figure1}
\begin{center}
\includegraphics*[width=9cm,angle=0]{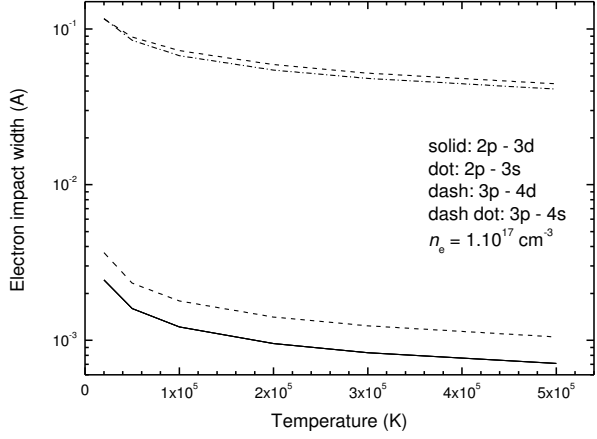}
\end{center}
\caption{Temperature dependence of the electron impact width for B IV $1s2p ^{1}P^{o} - 1s3s ^{1}S, 1s2p ^{1}P^{o} - 1s3d ^{1}D,
1s3p ^{1}P^{o} - 1s4s ^{1}S$ and $1s3p ^{1}P^{o} - 1s4d ^{1}D$ transitions at an electron density of $1.10^{17}$ cm$^{-3}$.}
\end{figure}

\begin{figure}
\label{figure2}
\begin{center}
\includegraphics*[width=9cm,angle=0]{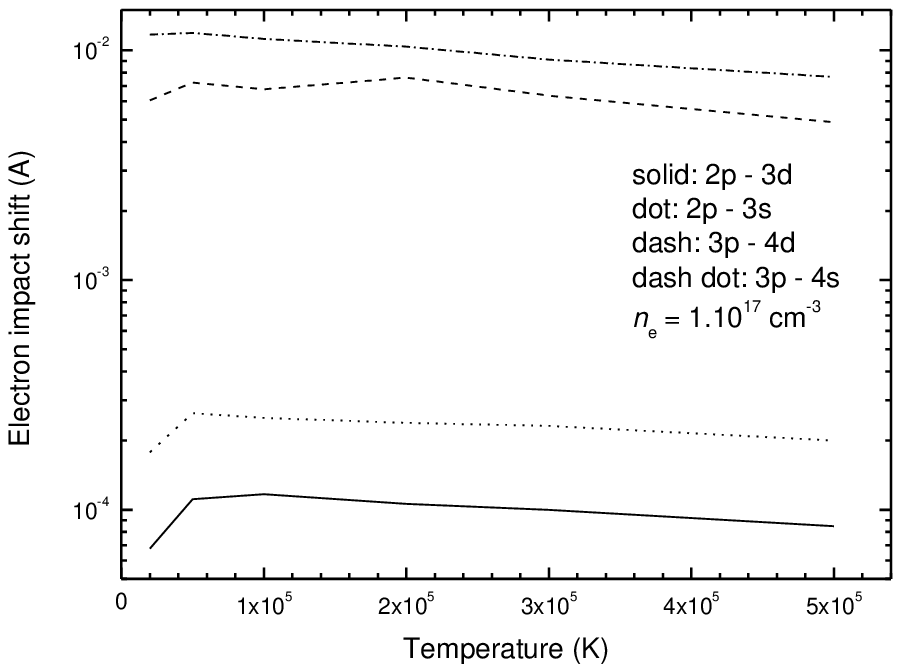}
\end{center}
\caption{Temperature dependence of the electron impact shift for $1s2p ^{1}P^{o} - 1s3s ^{1}S, 1s2p ^{1}P^{o} - 1s3d ^{1}D,
1s3p ^{1}P^{o} - 1s4s ^{1}S$ and $1s3p ^{1}P^{o} - 1s4d ^{1}D$ transitions at an electron density of $1.10^{17}$ cm$^{-3}$.}
\end{figure}

\begin{figure}
\label{figure3}
\begin{center}
\includegraphics*[width=8cm]{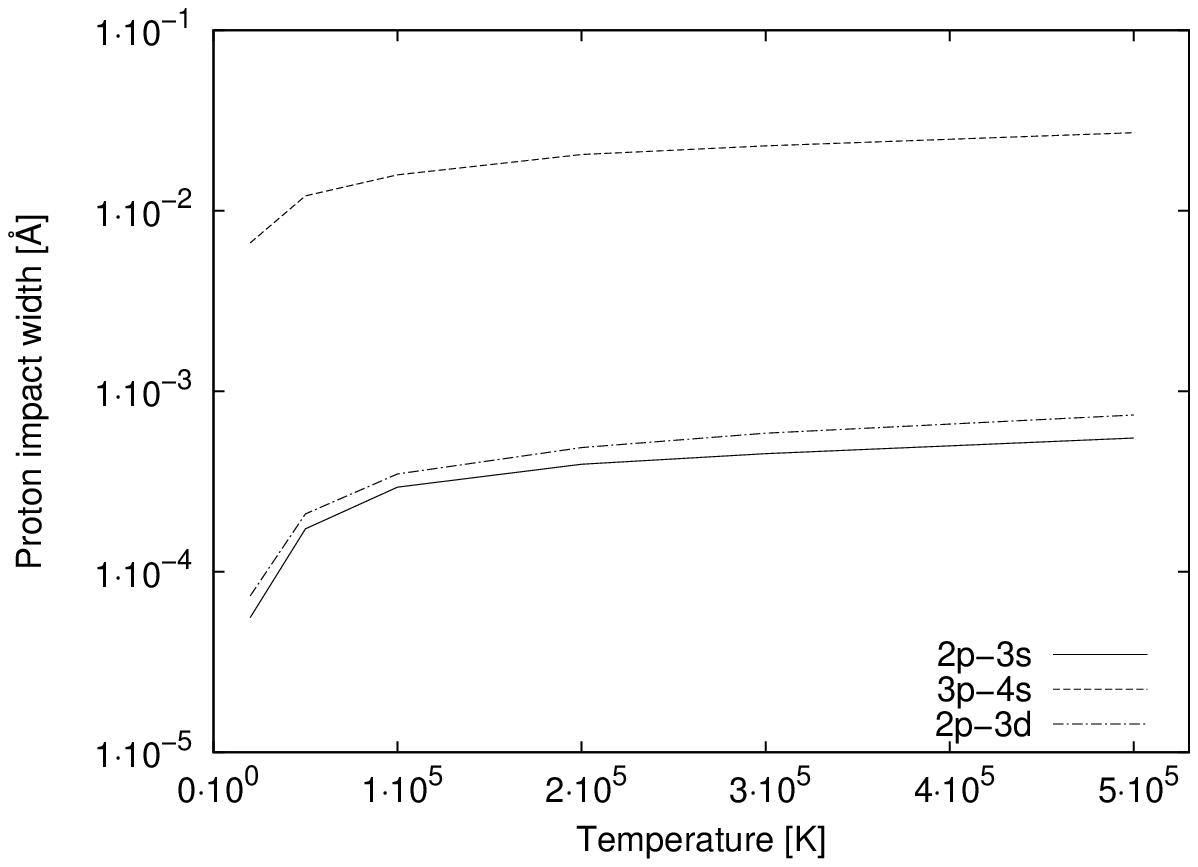}
\end{center}
\caption{Temperature dependence of the proton impact width for $1s2p ^{1}P^{o} - 1s3s ^{1}S, 1s2p ^{1}P^{o} - 1s3d ^{1}D,
1s3p ^{1}P^{o} - 1s4s ^{1}S$ and $1s3p ^{1}P^{o} - 1s4d ^{1}D$ transitions at an electron density of $1.10^{17}$ cm$^{-3}$.}
\end{figure}

\begin{figure}
\label{figure4}
\begin{center}
\includegraphics*[width=8cm]{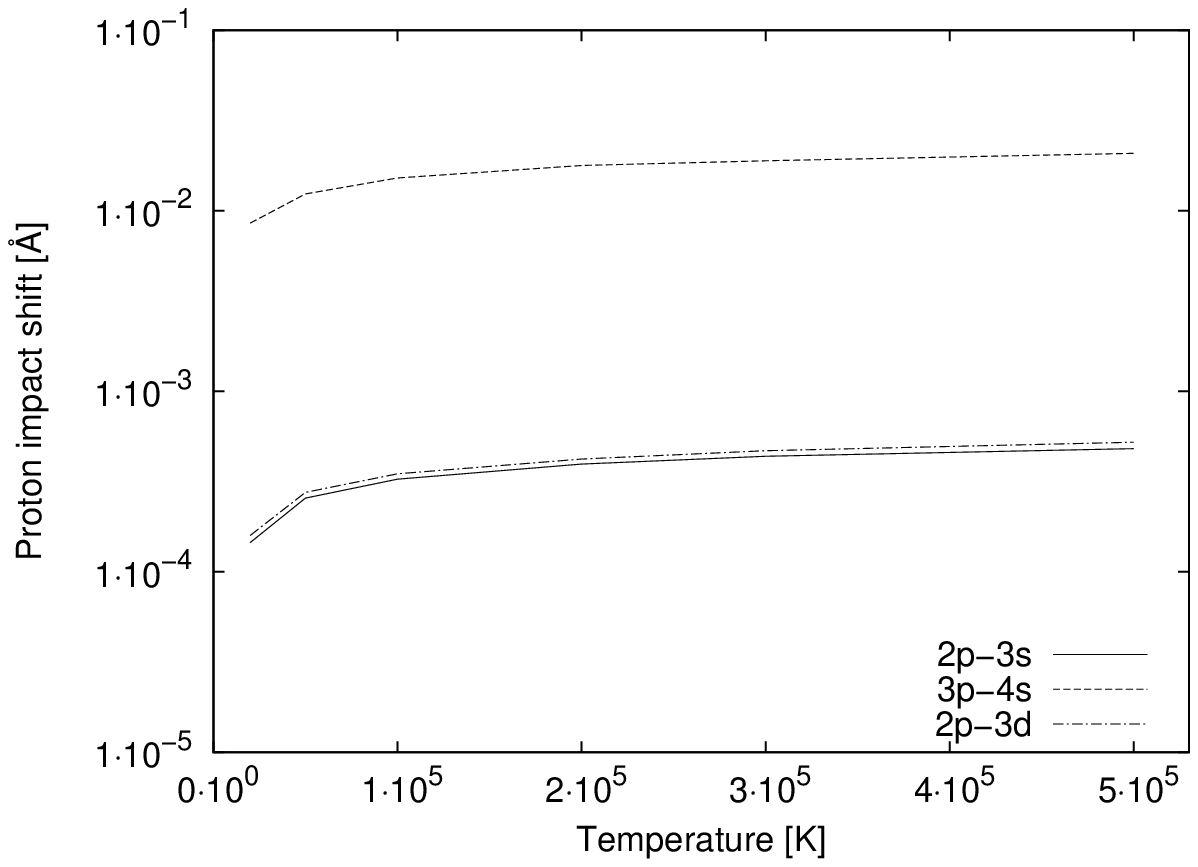}
\end{center}
\caption{Temperature dependence of the proton impact shift for $1s2p ^{1}P^{o} - 1s3s ^{1}S,
1s2p ^{1}P^{o} - 1s3d ^{1}D, 1s3p ^{1}P^{o} - 1s4s ^{1}S$ and $1s3p ^{1}P^{o} - 1s4d ^{1}D$
transitions at an electron density of $1.10^{17}$ cm$^{-3}$.}
\end{figure}

\begin{figure}
\label{figure5}
\begin{center}
\includegraphics*[width=9cm,angle=0]{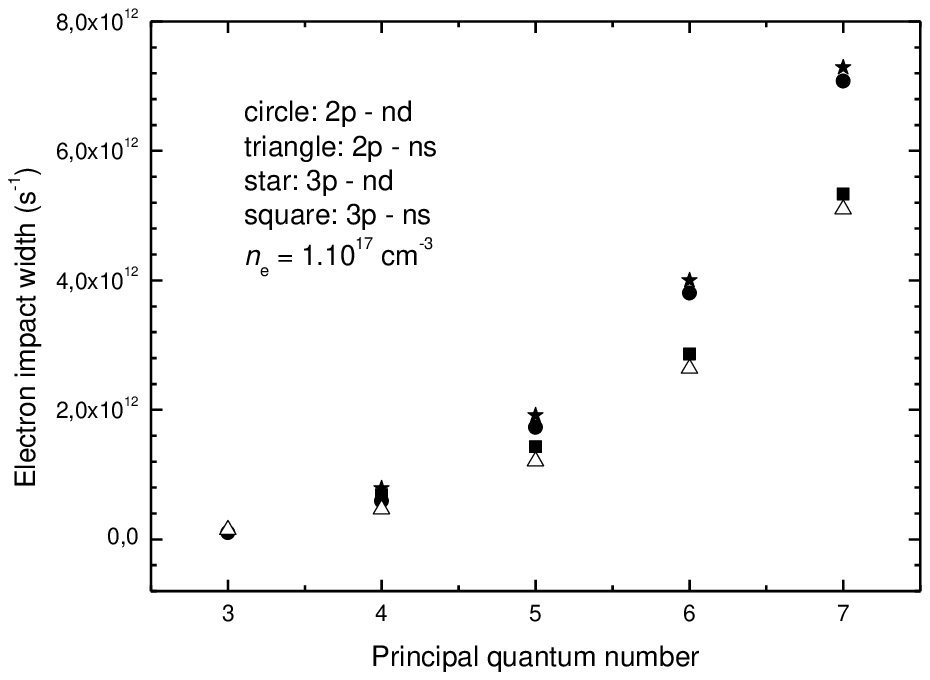}
\end{center}
\caption{Principal quantum number dependence of the electron-impact width for transitions within
$1s2p ^{1}P^{o} - 1sns ^{1}S, 1s2p ^{1}P^{o} - 1snd ^{1}D, 1s3p ^{1}P^{o} - 1sns ^{1}S$ and
$1s3p ^{1}P^{o} - 1snd ^{1}D$ spectral series
at an electron density of $1.10^{17}$ cm$^{-3}$ and T=200 000 K. }
\end{figure}

\begin{figure}
\label{figure6}
\begin{center}
\includegraphics*[width=9cm,angle=0]{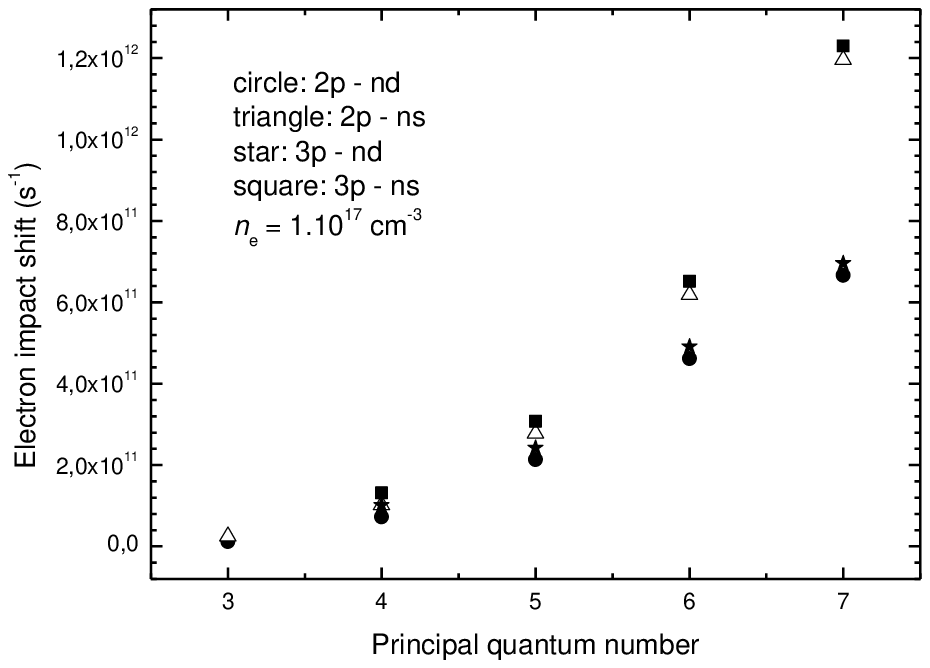}
\end{center}
\caption{Principal quantum number dependence of the electron-impact width for transitions within
$1s2p ^{1}P^{o} - 1sns ^{1}S, 1s2p ^{1}P^{o} - 1snd ^{1}D, 1s3p ^{1}P^{o} - 1sns ^{1}S$ and
$1s3p ^{1}P^{o} - 1snd ^{1}D$ spectral series
at an electron density of $1.10^{17}$ cm$^{-3}$ and T=200 000 K. }
\end{figure}

\end{twocolumn}

\end{document}